\documentclass[aps,pra,showpacs,amsmath,amssymb,amsfonts,twocolumn,superscriptaddress,floatfix,nofootinbib]{revtex4}
\usepackage{psfig}
\usepackage{graphicx}

\newcommand{\eqname}[1]{\label{eq:#1}}
\newcommand{\bgar}{\begin{eqnarray}}
\newcommand{\enar}[1]{\label{eq:#1}\end{eqnarray}}
\newcommand{\valass}[1]{\left|#1\right|}

\newcommand{\x}{ {\bf x}}

\newcommand{\xperp}{ {\bf x}_\perp}

\newcommand{\eq}[1]{(\ref{eq:#1})}

\begin{document}

\title{Nonlinear atom optics and bright gap soliton generation in
finite optical lattices}

\affiliation{Laboratoire Kastler Brossel, \'Ecole Normale
Sup\'erieure, 24 rue Lhomond, 75231 Paris Cedex 05, France}
\affiliation{Dipartimento di Chimica Fisica ed Inorganica, 
Viale del Risorgimento 4, I-40136 Bologna, Italy}
\affiliation{INFM, Dipartimento di Fisica, Universit\`a di Pisa, I-56126 Pisa, Italy} 
\affiliation{Scuola Normale Superiore, Piazza dei Cavalieri
7, I-56126 Pisa, Italy}
\affiliation{INFM, Scuola Normale Superiore, Piazza dei Cavalieri
7, I-56126 Pisa, Italy}

\author{Iacopo Carusotto}
\email{Iacopo.Carusotto@lkb.ens.fr}
\affiliation{Laboratoire Kastler Brossel, \'Ecole Normale
Sup\'erieure, 24 rue Lhomond, 75231 Paris Cedex 05, France}
\affiliation{INFM, Scuola Normale Superiore, Piazza dei Cavalieri
7, I-56126 Pisa, Italy}

\author{Davide Embriaco}
\affiliation{Dipartimento di Chimica Fisica ed Inorganica, 
Viale del Risorgimento 4, I-40136 Bologna, Italy}
\affiliation{INFM, Dipartimento di Fisica, Universit\`a di Pisa, I-56126 Pisa, Italy} 

\author{Giuseppe C. La Rocca}
\affiliation{Scuola Normale Superiore, Piazza dei Cavalieri
7, I-56126 Pisa, Italy}
\affiliation{INFM, Scuola Normale Superiore, Piazza dei Cavalieri
7, I-56126 Pisa, Italy}

\begin{abstract}
We theoretically investigate the transmission dynamics of coherent
matter wave pulses across finite optical lattices in both the linear and
the nonlinear regimes. The shape and the intensity of the transmitted pulse
 are found to strongly depend on
the parameters of the incident pulse, in particular its velocity and
density: a clear physical picture for the main features observed in the
numerical simulations is given in terms of the atomic band dispersion in the 
periodic potential of the optical lattice. 
Signatures of nonlinear effects due the atom-atom interaction are
discussed in detail, such as atom optical limiting and atom optical
bistability.
For positive scattering
lengths, matter waves propagating close to the top of the valence band
are shown to be subject to modulational instability. 
A new scheme for
the experimental generation of narrow bright gap solitons from a wide
Bose-Einstein condensate is proposed: the modulational instability 
is seeded in a controlled way starting from the strongly modulated
density profile of a standing matter wave and the solitonic nature of
the generated pulses is checked from their shape and their collisional 
properties. 
\end{abstract}



\date{\today}

\maketitle



In recent years, a great interest has been devoted to theoretical
as well as experimental studies of the propagation of matter waves
in the periodic potential of optical lattices. The first experiments
were carried out using ultracold atomic
samples~\cite{ColdAtomsLatticeExp}; later on, the
realization of
atomic Bose-Einstein condensates (BECs)~\cite{BEC} and their coherent loading
into optical lattices~\cite{BECLatticeExp} have opened the possibility
of investigating
features which follow from the coherent nature of the Bose-condensed atomic
sample~\cite{BECGapSolit,InstabilityBEC}.

At the same time, the propagation of light waves in linear and nonlinear 
periodic dielectric structures has been a very active field of research: global
photonic band gaps (PBG) have been observed~\cite{PBG} and 
one-dimensional nonlinear periodic systems such
as nonlinear Bragg fibers~\cite{SterkeReview} are actually
under intense investigation given the wealth of different
phenomena including optical bistability, modulational
instability and solitonic propagation that can be
observed~\cite{BECLatticeTh,GapSolitExp,SelfPulsing}.

Given the very close analogy between the behavior of coherent matter
waves and nonlinear optics, we expect that the concepts currently
used to study the physics of nonlinear Bragg fibers can be fruitfully
extended to the physics of coherent matter waves in optical lattices: the optical potential of the
lattice plays in fact the role of the periodic refractive index, the atom-atom
interactions are the atom-optical analog of a Kerr-like nonlinear
refractive index and the Gross-Pitaevskii equation of mean-field
theory corresponds to Maxwell's equation with a nonlinear polarization
term~\cite{NLO}.

In the present paper, we shall report a theoretical investigation of 
the transmission dynamics of coherent matter pulses 
which incide onto a finite optical
lattice with a velocity of the order of the Bragg
velocity. In this
velocity range, Bragg reflection processes are most
effective and the atomic dispersion in the lattice is completely 
different from the
free-space one.
Depending on the value of the density, spatial size and velocity of
the incident atomic cloud, as well as on the depth and length 
of the lattice, a number of different
behaviors are predicted by numerical simulations; here, 
we shall focus our attention on the shape of the transmitted pulse
just after it has crossed the lattice as well as while it is still
propagating in the lattice.
In particular, we shall discuss 
a new mechanism which can be used to generate narrow bright atomic gap
solitons propagating along the lattice starting from 
a wide incident Bose-Einstein condensate.
For more details on the continuous-wave transmission and reflection 
spectra at linear regime, the reader can refer
to~\cite{Ozeri,Santos,Santos2,Santos3,AtomFP}; some aspects of the
linear pulse dynamics are discussed in~\cite{Santos3}.

The geometry considered in the present paper as well as
in~\cite{Ozeri,Santos,Santos2,Santos3,AtomFP} is significantly different from
the one usually
considered in recent experimental works~\cite{BECLatticeExp} of BEC
dynamics in optical lattices in which an optical lattice is
switched on and superimposed to a stationary condensate; the dynamics of the
condensate inside the lattice
is then studied in response to some external force such as gravity, an
acceleration of the
lattice, or a spatial translation of the magnetic
potential.

The present paper is organized as follows: sec.\ref{sec:System} describes
the physical system under examination. 
In sec.\ref{sec:Bands} we review some basic concepts on the dispersion of
matter waves in the periodic potential of an infinite optical lattice and
present some simple analytical calculations which accurately reproduce the
dispersion of the atomic bands in the neighborhood of the first
forbidden gap.
The linear regime propagation of coherent matter wave pulses across
finite lattices is the subject of sec.\ref{sec:Linear}. The intensity
as well as the shape of the transmitted pulse are found to strongly
depend on the properties of the weak incident pulse, in particular its
velocity and spatial size; a simple interpretation of the observed
phenomena in terms of allowed bands and forbidden gaps is provided.
In sec.\ref{sec:Nonlinear}, we discuss the effect of atom-atom
interactions on the propagation of the pulse in the different cases: 
simple explanations for the observed behavior are put forward 
in terms of familiar concepts of nonlinear optics such as optical
limiting, optical bistability, or modulational instability; in
particular, modulational instability is shown to occur for positive
scattering length atoms when the matter wave is propagating at the top 
of the valence band; at this point of the dispersion, the effective
mass is in fact negative.
In sec.\ref{sec:Soliton}, we propose a new scheme to control the
modulational instability of valence band atoms in order to obtain a
narrow bright gap soliton from a wide condensate: as an initial seed for the
instability, the standing matter wave pattern created by the
interference of the incident and reflected waves is used. The
solitonic nature of the generated pulses is verified by looking at
their dynamical and collisional properties as well as by comparing the
pulse shape with the analytical predictions of the envelope-function
approximation discussed in sec.\ref{sec:SolitAnal}.
Conclusions are finally drawn in sec.\ref{sec:Conclu}.

\section{The physical system}
\label{sec:System}

We consider a Bose-condensed atomic cloud in a quasi-1D geometry
in which the transverse motion is frozen by the confining potential of
an optical or magnetic atomic waveguide~\cite{waveguide}.
Gravity is made immaterial either by placing the waveguide axis along
the horizontal plane or by counterbalancing the gravitational field
with a suitable magnetic field gradient.

\begin{figure}[htbp]
\includegraphics[width=3in]{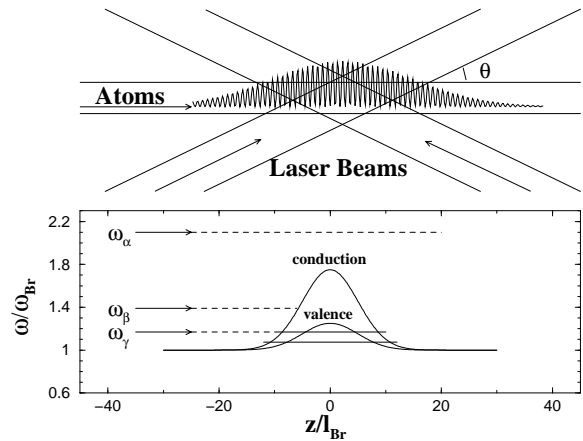}
\caption{
Upper panel: schematic plot of the experimental set-up under
consideration.
Lower panel: spatial dependence of the local band edge energies.
\label{fig:Setup}}
\end{figure}

A periodic potential is created along the waveguide axis by means
of a pair of far-off resonance laser beams of frequency $\omega_L$ and
wave vector $k_L=\omega_L/c$ crossing the waveguide at an angle
$\theta$ as in fig.\ref{fig:Setup}~\cite{BECLatticeExp}: denoting with
\mbox{$\Omega_L(z)=|{\vec d}\cdot
{\vec E(z)}|/\hbar$} the (slowly varying) single beam Rabi
frequency and with $\omega_{at}$ the atomic transition frequency,
the optical potential experienced by the atoms is given by $V_{\rm
opt}(z)=V_0(z)\cos^2k_{\rm Br}z$, with 
$V_0(z)=\hbar\Omega_L(z)^2/(\omega_L-\omega_{at})$ and $k_{\rm
Br}=k_L\cos\theta$.
For a red- or blue-detuned laser field, the optical potential is respectively 
attractive or repulsive; the lattice period $\ell_{\rm Br}=\pi/k_{\rm
Br}$ can be tuned by varying the angle $\theta$.
The longitudinal envelope  $V_0(z)$ of the lattice potential is determined by the profile
of the laser beam waist and is assumed to smoothly vary on a length
scale significantly longer than the lattice period $\ell_{\rm Br}$; unless
differently specified, $V_0(z)$ will be taken as a
Gaussian
\begin{equation}
V_0(z)=V_0\,e^{-z^2/2 w_l^2}
\end{equation}
of height $V_0$ and spatial length $w_l$.

If both the kinetic and the interaction energy are much smaller than
the transverse level spacing of the waveguide, this latter can be
considered as being a single-mode one and the condensate wave function can be
written in the factorized form 
\begin{equation}
\eqname{psiFactor}
\psi(\x)=\psi(z)\psi_\perp(\xperp)
\end{equation} where
$\psi_\perp(\xperp)$ is the ground-state eigenfunction of the 
transverse confining potential with the appropriate
$\int\!d^2\x_\perp\,|\psi_\perp(\x_\perp)|^2=1$ normalization.
Within this approximation, the dynamics
of the condensed atomic cloud can be described by a one-dimensional
Gross-Pitaevskii equation
\begin{multline}
i\hbar\frac{\partial \psi(z,t)}{\partial
t}=\left(-\frac{\hbar^2}{2m_0}\frac{\partial^2}{\partial
z^2}+V_{\rm opt}(z)+\right. \\ \left.+g_{\rm 1D}\valass{\psi(z,t)}^2\right)\psi(z,t)
\eqname{GP}
\end{multline}
where $m_0$ is the atomic mass and the renormalized 1D effective
interaction $g_{\rm 1D}$ is written in terms of
the usual 3D scattering length $a$ as~\cite{BECSolitTh}
\begin{equation}
g_{\rm 1D}=\frac{4\pi\hbar^2 a}{m_0}\int\!d^2\xperp\,\valass{\psi_\perp(\xperp)}^4.
\eqname{g1D}
\end{equation}

\section{Allowed bands and forbidden gaps}
\label{sec:Bands}
As it happens to electrons in crystalline
solids~\cite{SolidState} and light in periodic dielectric systems such as
photonic band gap crystals~\cite{PBG}, the atomic
dispersion in an infinite periodic potential is characterized at linear
regime (i.e. in the non-interacting case) by allowed bands and forbidden gaps.

When the depth $V_0$ of the lattice potential is weak with respect to the
Bragg energy $\hbar\omega_{\rm Br}=\hbar^2 k_{\rm Br}^2/2m_0$ the lowest part of the
band structure can be accurately described within a {\em nearly-free atom}
approximation~\cite{AtomFP} in which only two coupled
modes~\cite{SterkeReview} are taken into
account
\begin{equation}
\psi(z)=a_{\rm f}e^{ikz}+a_{\rm b}e^{i(k-2k_{\rm Br})z}.
\eqname{TwoModes}
\end{equation}
In this restricted $({\rm f},{\rm b})$ basis, the linear regime Hamiltonian has the following form
\begin{equation}
{\rm H}=\left(
\begin{array}{cc}
\frac{\hbar^2 k^2}{2m_0}+\frac{V_0}{2} & \frac{V_0}{4} \\
\frac{V_0}{4} & \frac{\hbar^2(k-2k_{\rm Br})^2}{2m_0}+\frac{V_0}{2}.
\end{array}\right)
\end{equation}
and the eigenenergies $\hbar\omega_\pm$ are equal to
\begin{multline}
\hbar\omega_{\pm}(k)=\hbar\omega_{\rm Br}+\frac{V_0}{2}+\\ +\hbar\omega_{\rm Br}\left[\left(\frac{\Delta k}{k_{\rm Br}}\right)^2\pm
2\sqrt{\left(\frac{\Delta k}{k_{\rm
Br}}\right)^2+\left(\frac{V_0}{8\hbar\omega_{\rm Br}}\right)^2}\right]
\eqname{Bands}
\end{multline}
where we have set $\Delta k=k-k_{\rm Br}$; the positive sign holds for
the upper, {\em conduction band} and the
minus sign for the lower, {\em valence band}.
No states are present between $\omega_{\rm Br}+V_0/4$ and
$\omega_{\rm Br}+ 3V_0/4$: this is the lowest {\em forbidden gap} in
which the matter waves can not propagate through the lattice.

The group velocity is given by
\begin{multline}
\eqname{vg}
v_{\rm g}^\pm(k)=\frac{\partial \omega_\pm}{\partial k}=\\ =\frac{v_{\rm
Br}\,\Delta k}{k_{\rm Br}}\left\{1\pm\left[\left(\frac{\Delta k}{k_{\rm
Br}}\right)^2+\left(\frac{V_0}{8\hbar\omega_{\rm Br}}\right)^2\right]^{-1/2}\right\};
\end{multline}
close to band edge ($\hbar v_{\rm Br}\,\Delta k\ll V_0$), the group velocity $v_{\rm
g}^\pm$ is much reduced with respect to the Bragg velocity $v_{\rm
Br}=\hbar k_{\rm Br}/m_0$; further away  ($\hbar v_{\rm Br}\,\Delta
k\gg V_0$), it recovers the free-space value $\hbar k/m$.

The effective mass of the atom is given by
\begin{multline}
\eqname{meff}
\frac{1}{m_{\rm eff}(\Delta k)}=\frac{1}{\hbar}\frac{\partial
^2 \omega_{\pm}}{\partial k^2}=\\ =\frac{1}{m_0}\left\{1\pm\frac{8\hbar\omega_{\rm
Br}}{V_0}\left[1+\left(\frac{\Delta k/k_{\rm Br}}{V_0/8\hbar
\omega_{\rm Br}}\right)^2\right]^{-3/2}\right\}.
\end{multline}
At band edge ($\Delta k=0$), $m_{\rm eff}$ is much
smaller in absolute value than the free-space atomic mass $m_0$; as
usual, it is positive at the conduction band edge, while
it is negative at the valence band edge.
Further away from the band edge, 
$m_{\rm eff}$ coincides with the positive free-space
value $m_0$~\cite{SolidState}.

Around the band edge, the weights of the forward and
backward traveling waves are comparable
$|a_{\rm f}|\simeq|a_{\rm b}|$ for both valence and conduction
bands: the density profile of the Bloch eigenfunction thus has a
standing wave shape with a spatial period equal to the lattice period 
$\ell_{\rm Br}$.
Further away from the gap one component dominates the other:
the Bloch eigenfunction has then a running wave character and the density
profile is uniform over the unit cell of the lattice.

\section{Propagation in the linear regime}
\label{sec:Linear}
We now consider a coherent matter pulse (e.g. extracted from a Bose-Einstein
condensate) which is moving along the waveguide with a uniform velocity $v_0$
close to the Bragg velocity $v_{\rm Br}$. 
Initially the atomic pulse is far outside the lattice.
The initial density distribution of atoms in the cloud is taken as
having a Gaussian shape
\begin{equation}
\psi_{\rm inc}(z)=\psi_{\rm max}\, e^{ik_0 z}\, e^{-(z-z_0)^2/2 w_0^2};
\end{equation}
the carrier momentum is $\hbar k_0=m_0 v_0$ and the carrier kinetic
energy $\hbar\omega_0=\hbar^2 k_0^2/2 m_0$; since the wave packet is finite in
space, its Fourier transform ${\tilde \psi}(k)$ has a finite momentum 
spread $\sigma_k=1/w_0$ around $k_0$.

\begin{figure}[htbp]
\includegraphics[width=3in]{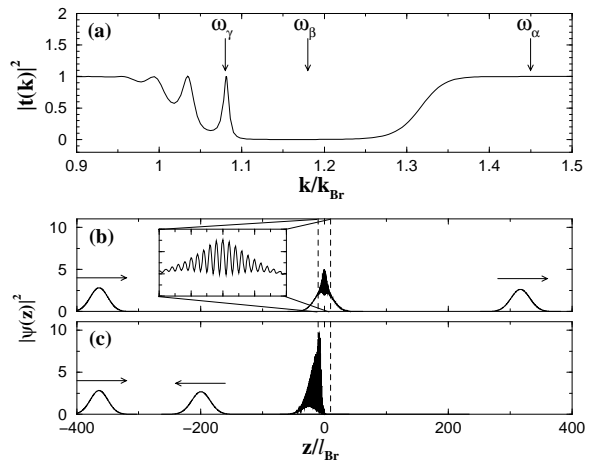}
\caption{
Upper (a) panel: linear transmission spectrum across a repulsive
($V_0/\omega_{\rm Br}=1$) optical lattice with Gaussian profile
($w_l/\ell_{\rm Br}=5$). 
Lower panels: linear regime time-evolution of Gaussian incident 
pulses ($w_0/\ell_{\rm
Br}=20$) centered at respectively $\omega_0=\omega_\alpha$ (b) (transmitted) and
$\omega_0=\omega_{\beta}$ (c) (reflected). The spatial extension
of the lattice is indicated by the vertical dashed lines.
\label{fig:Linear}}
\end{figure}

If the density is sufficiently low and the interactions sufficiently
weak, the nonlinear term in the Gross-Pitaevskii equation
\eq{GP} can be neglected and the time-evolution of the matter field
is accurately described by a linear Schr\"odinger equation. 
In this case, the superposition principle holds and the 
transmission of the pulse can be well described in momentum space 
in terms of the wave vector-dependent transmission
amplitude $t(k)$ 
\begin{equation}
{\tilde \psi}_{\rm t}(k)=t(k)\,{\tilde \psi}_{\rm inc}(k).
\end{equation}
If the whole of the incident wave packet is contained in either a
transmitting or a reflecting region of the spectrum
(fig.\ref{fig:Linear}a), it will be respectively transmitted
(fig.\ref{fig:Linear}b) or reflected
(fig.\ref{fig:Linear}c) without sensible reshaping. 
As it has been discussed in full detail in previous papers~\cite{AtomFP}, complete transmission 
occurs whenever propagating states are available at all spatial positions
for the frequencies of interest. 
Given the smooth envelope of the
lattice, interface reflections do not occur and the matter wave
 adiabatically follows the shape of the corresponding
Bloch state; the typical sinusoidal shape of band edge Bloch
wave functions~\cite{PBG,SolidState} multiplying the broad pulse
envelope can be clearly recognized while the wave packet is crossing the
lattice (fig.\ref{fig:Linear}b, inset).
On the other hand, if the wave packet frequency falls inside the
forbidden gap at some spatial positions, the matter wave is not
able to cross that region and is then nearly completely reflected back (fig.\ref{fig:Linear}c); the inversion point
is located at the beginning of the forbidden region, i.e. at the point
at which the carrier frequency $\omega_0$ enters the forbidden gap.

\begin{figure}[htbp]
\includegraphics[width=3in]{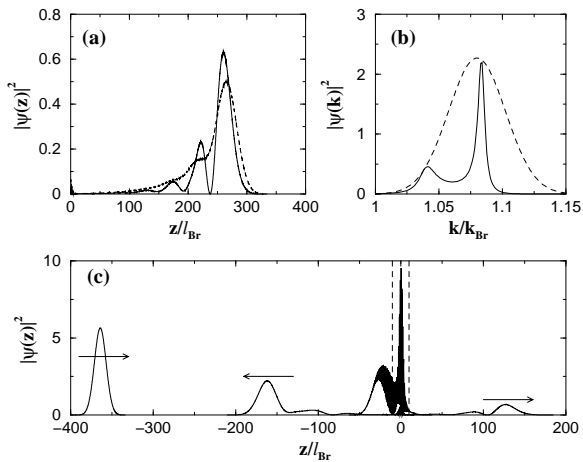}
\caption{Lower panel (c): time-evolution of a weak and short
($w_0/\ell_{\rm Br}=10$) Gaussian pulse
close to resonance with a
quasi-bound mode at $\omega_\gamma\simeq 1.17\,\omega_{\rm Br}$ (see fig.\ref{fig:Linear}a); lattice parameters as in
figs.\ref{fig:Setup} and \ref{fig:Linear}. The spatial extension
of the lattice ($w_l/\ell_{\rm Br}=5$) is indicated by the vertical dashed lines.
Upper panels: (a) expanded view of the transmitted
pulse shape for the pulse in (c) (solid line) and for a longer
$w_0/\ell_{\rm Br}=20$ pulse
(dashed line); (b) incident (dashed) and transmitted (solid) spectra
for the incident pulse in (c). 
\label{fig:LinearRes}}
\end{figure}

When one or more resonance peaks due to quasi-bound modes\footnote{In the analogy with optics, these quasi-bound modes
correspond to the resonance peaks of a Fabry-Perot interferometer or, more
closely, to the cavity modes of a DBR
microcavity~\cite{DBRMicrocav}. A more detailed discussion about
quasi-bound modes in optical lattices can be found in~\cite{AtomFP}
and in~\cite{Santos2,Santos3}.}
are contained within the spectrum of the
incident wave packet, the pulse will be partially reflected and partially
transmitted (fig.\ref{fig:LinearRes}b-c); the strong frequency
dependence of the transmission amplitude leads to a profound reshaping
of the spatial profile of the pulse (fig.\ref{fig:LinearRes}a).
An incident pulse of linewidth much wider than the quasi-bound mode
linewidth has in fact a time duration much shorter than the characteristic
decay time of the mode; this latter can therefore be considered as being
impulsively excited by the incident pulse and than exponentially
decaying on a much longer time scale.
The exponential tail shown by the transmitted pulse profile when a
single quasi-bound mode is excited (dashed line in
fig.\ref{fig:LinearRes}a) is a clear signature of transmission
occurring via a single resonant quasi-bound mode~\cite{CCT4}.

If several isolated modes are instead present in the frequency interval
encompassed by the incident spectrum, the spectrum of the transmitted
pulse will contain several peaks (fig.\ref{fig:LinearRes}c) and a complex shape will result
from the interference of the different frequency components.
For example, if the incident pulse has a duration comparable to the
dephasing time of a pair of neighboring modes (i.e. the inverse of
their frequency difference), both of them will be impulsively excited and
the profile of the transmitted pulse will show oscillations following 
the time evolution of the relative phase (solid line in
fig.\ref{fig:LinearRes}a); these oscillations can be
interpreted as {\em quantum beatings} between the two coherently excited
quasi-bound modes coupled to the same continuum of propagating modes 
outside the lattice\footnote{Similar oscillations have been studied
in~\cite{Santos3} in the case in which a large number of closely
spaced quasi-bound modes are excited.}

\section{Nonlinear regime and modulational instability}
\label{sec:Nonlinear}
For higher values of the atomic density and the coupling constant $g_{\rm
1D}$, the effect of the atom-atom interactions is no longer negligible
 and the mean-field nonlinear term $g_{\rm 1D}|\psi|^2$ in \eq{GP} 
starts playing an important role in the dynamics; 
in the following, we shall focus on the experimentally most
relevant case of a positive scattering length $a>0$, which means a repulsive
interaction among atoms in free space.

In~\cite{AtomFP} we discussed the case of a continuous wave (cw) beam of
coherent atoms incident on a finite lattice; depending on the detuning of
the incident beam with respect to the frequency of a quasi-bound mode
of the lattice, atom optical limiting or bistability was predicted 
for an incident beam respectively on or above the resonance frequency.
Here we shall consider the more realistic case of a finite atomic wave packet 
incident on a finite optical lattice; its central frequency
$\omega_0$ is taken to be close to a quasi-bound mode of the lattice
and its temporal duration much longer than the lifetime of the mode so
that the frequency spread is narrower than the mode linewidth.

\begin{figure}[htbp]
\includegraphics[width=3in]{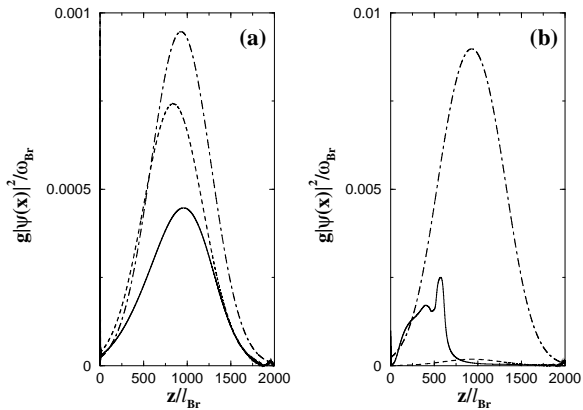}
\caption{Transmitted pulse shape for Gaussian
incident pulses on resonance (a) and $0.04\,\omega_{\rm Br}$ 
above (b) resonance with the quasi-bound mode of the lattice at
$\omega_{\gamma}\simeq 1.17\,\omega_{\rm Br}$ (solid lines). For
comparison, same calculations neglecting the interaction term (dashed
lines) and in the absence of the lattice (dot-dashed lines).
Same lattice parameters as in figs.\ref{fig:Linear} and
\ref{fig:LinearRes}; the pulse starts at $z_0/\ell_{\rm Br}=-1000$
with a Gaussian width equal to $w_0/\ell_{\rm Br}=480$.
\label{fig:Nonlinear}}
\end{figure}

\begin{figure}[htbp]
\includegraphics[width=3in]{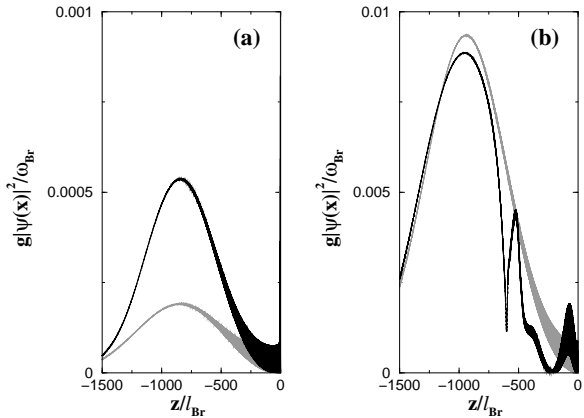}
\caption{Reflected pulse shapes for the same values of the physical
parameters as in fig.\ref{fig:Nonlinear} (black lines). For
comparison, same calculation neglecting the interaction term (grey lines).
\label{fig:NonlRefl}}
\end{figure}

First we consider the case of an incident wave packet with a
center-of-mass kinetic energy $\hbar\omega_0$ exactly on
resonance with a quasi-bound mode (fig.\ref{fig:Nonlinear}a): 
in the absence of interactions (dashed line), the wave packet is
nearly completely transmitted without any shape distortion; only a
small fraction of the pulse is reflected (fig.\ref{fig:NonlRefl}a).
Thanks to the resonance condition with the quasi-bound mode, the
atomic density inside the lattice is significantly larger than the
one in the incident wave packet.
In the presence of interaction, the main effect of nonlinearity is to 
blue-shift the quasi-bound mode
frequency~\cite{NLO} and push it out of
resonance with the incident beam; this negative feedback effect,
already present in the cw treatment~\cite{AtomFP}, gives not only a saturation of the
transmission as a function of the incident density, but also a
significant reshaping of the wave packet: the higher
density part is cut and the resulting wave packet is flattened.
The negative feedback in the transmission obviously corresponds to a
positive feedback in the reflection.

In the opposite case of a wave packet incident with a kinetic energy
higher than the quasi-mode frequency, a low density wave packet is
nearly completely reflected (fig.\ref{fig:Nonlinear}b, dashed line).
At higher densities, the cw calculations in~\cite{AtomFP} have
predicted the appearance of atom optical bistability effects: several
stationary states with different transmitted intensities can indeed be found
for the same value of the incident intensity.
On such grounds, we are able to put forward a physical interpretation of
the transmitted pulse shape in the presence of interactions 
shown as a solid line in
fig.\ref{fig:Nonlinear}b: only a small fraction of the leading
part of the pulse is transmitted since the incident frequency is out
of resonance with the empty quasi-bound mode; moreover, this part of
the incident
wave packet has been accelerated by the
repulsive mean-field interactions before reaching the lattice and thus
pushed even further off resonance.
For the same reason, the trailing part of the pulse
has been instead slowed down with respect to the central velocity and thus 
results closer to resonance to the quasi-bound mode; when this trailing
part of the pulse reaches the lattice, the quasi-bound mode starts
to be effectively populated.
The main effect of the interactions among the atoms 
in the quasi-bound mode is to push its frequency closer to resonance with the
incident wave packet: the sudden increase in the density of the
transmitted pulse which is apparent in fig.\ref{fig:Nonlinear}b is a
direct consequence of this positive feedback. This behaviour is analog to the
jump from
the lower to the upper branch of the bistability loop 
which occur in the cw case for incident intensities growing beyond the switch-on threshold.
The shape of the reflected pulse is complementary to the one of the
transmitted one: in the presence of the nonlinearity, the reflected
pulse shows a dip corresponding to the switch-on of the transmission (fig.\ref{fig:NonlRefl}b).

Provided the interaction energy is much smaller than
the spacing of the different quasi-bound modes, the transmission
dynamics is mostly determined by a single resonant
mode and the shape of the matter field inside the lattice is
fixed by the eigenfunction of the mode; this guarantees that no {\em modulational
instability}~\cite{InstabilityBEC,ModInstSuppr} 
can take place\footnote{From a different point of view, this suppression of the
modulational instability can be interpreted in the following terms: if
the spatial extension of the condensate wave packet is small enough,
the excitation of the long wavelength modes which are 
responsible for the modulational instability can not occur because of
the finite size of the system; this effect is well known from the
physics of trapped BECs with attractive interactions, which are stable
provided the number of atoms is sufficiently small for the effective
healing length to be larger than the condensate size~\cite{AttractiveBEC}.}
 even in the presence of an effectively attractive
interaction such as the one which occurs in the case of negative mass $m_{\rm
eff}<0$ valence band atoms for which the sign of the effective
interaction is reversed with respect to free space.

\begin{figure}[htbp]
\includegraphics[width=3in]{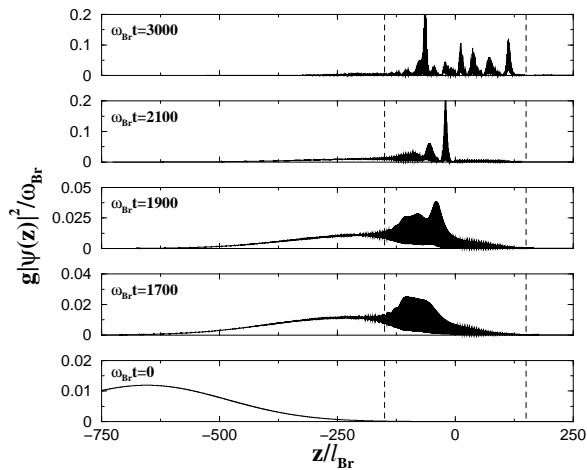}
\caption{Onset of modulational instability: propagation of a pulse of
valence band atoms ($\omega_0/\omega_{\rm
Br}=0.68$, $w_0/\ell_{\rm Br}=240$, $z_0/\ell_{\rm Br}=-650$) across an attractive lattice ($V_0/\omega_{\rm
Br}=-0.4$). The lattice has a flat profile in the central
$|z/\ell_{\rm Br}|<100$ region and Gaussian ($w_l/\ell_{\rm Br}=50$)
wings;  its spatial extension is indicated by the vertical dashed lines.
\label{fig:ModInstab}}
\end{figure}

In the absence of spatial confinement, a
spatially extended
wave packet of coherent valence band atoms is instead subjected to modulational
instability: consider for example an attractive lattice and an
incident wave packet with a kinetic energy just below the lower edge
of the gap.
In the present paper we shall limit ourselves to the case of a
sufficiently weak nonlinearity $g_{\rm 1D}|\psi|^2\ll V_0$ in order
for the band structure of the atomic dispersion not to be washed out
by the interaction term.
Since propagating valence band states are available at all spatial
positions, the pulse is able to penetrate inside the lattice without
any reflection. As soon as the pulse is in the lattice, the
modulational instability sets in 
and the initially uniform
envelope starts being modulated with an amplitude which grows
exponentially in time; at the end, a train of short and intense pulses is
found (fig.\ref{fig:ModInstab}). The seed for this modulational instability is
automatically provided by the density modulations which inevitably
arise while the pulse is entering the non-uniform lattice.
In order to make the discussion the simplest, this has been 
assumed to have an uniform depth in its central part and wings at its
ends; after the
initial preparation phase, the pulses therefore 
propagate through an effectively
uniform system\footnote{In the presence of a slowly
varying modulation of the lattice parameters, no significant coupling of the
internal and external degrees of freedom of a solitonic pulse is
expected to occur if the characteristic length of the inhomogeneity is
much longer than the size of the pulse~\cite{SolitAdiab}.}.

A similar modulational instability is well-known to occur in nonlinear
Bragg fiber optics~\cite{SterkeReview}; a cw laser beam traveling in the
conduction (valence) photonic band of a Bragg lattice with a focusing
(defocusing) nonlinear refractive index is subjected to self-pulsation
and therefore converted into a train of pulses.
Since the nonlinear refraction index can be
interpreted as an effective photon-photon
interaction~\cite{SolitReview,PhotInter}, the self-pulsing effect is
easily interpreted as the modulational instability following the
presence of an effectively attractive interaction.

\section{Controlled bright gap soliton generation}
\label{sec:Soliton}
A regular train of short optical pulses with a well defined
period and intensity has been theoretically shown to be produced if
the modulational instability of a cw laser beam propagating in a
lattice is seeded with a weak periodic intensity modulation~\cite{SelfPulsing}.
The shape of each pulse is stable during
propagation since the group velocity dispersion is counterbalanced by
the optical nonlinearity. Following the literature, we
shall call these pulses {\em gap solitons}~\cite{SterkeReview}, even
if from a strictly mathematical point of view they could be
classified only as {\em solitary waves}, since the wave equation \eq{GP} in a periodic potential is not exactly
integrable~\cite{SolitReview} and collisions between two such
solitons do not exactly preserve the pulse shape~\cite{AcevesWabnitz};
as we shall see in the following, 
the expression {\em gap soliton} is however physically justified by
the fact that the pulse distortion following a collision is generally small.
Since the first observation of optical gap solitons in 1995, intense
experimental activity is actually in progress for the generation and
characterization of gap solitons and Bragg modulational instabilities
in optical fibers~\cite{GapSolitExp}.

In very recent years, solitonic excitations are beginning to be
investigated also in the context of nonlinear atom
optics~\cite{BECSolitTh}:
 dark solitons in the form of stable density dips in a otherwise
uniform condensate have been recently observed~\cite{BECSolitExp};
bright gap solitons~\cite{BECGapSolit} and modulational instabilities~\cite{InstabilityBEC} are actually under intense
investigation; despite the positive atom-atom
scattering length, such stable atomic pulses can exist inside an
optical lattice thanks to the negative effective mass of valence band atoms.
In the present section we present a new method to generate narrow
bright gap solitons
starting from a wide atomic condensate incident on an optical lattice

\begin{figure}[htbp]
\includegraphics[width=3in]{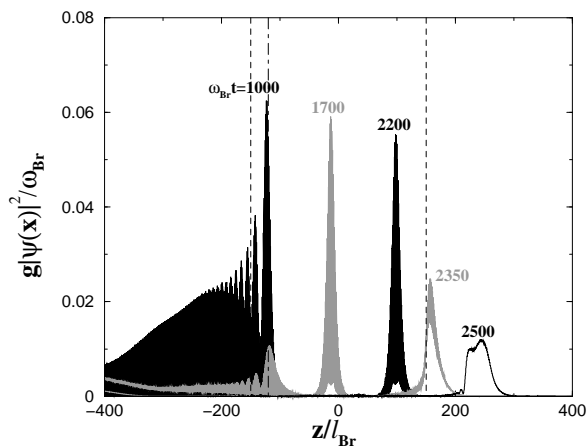}
\caption{
Controlled bright gap soliton generation from a standing matter wave
($\omega_0/\omega_{\rm Br}=0.72$, $w_0/\ell_{\rm Br}=240$, $z_0/\ell_{\rm
Br}=-650$; same lattice parameters as in fig.\ref{fig:ModInstab}): pulse
shape snapshots at different times.
Vertical dashed lines indicate the spatial extension of the lattice;
the dot-dashed line indicate the reflection point at linear regime.
\label{fig:Solit}}
\end{figure}

Consider a long coherent matter wave pulse (i.e. a Bose-condensed
atomic cloud) incident on the same attractive lattice as in the
previous section with a kinetic energy just above the lower edge
of the gap. 
At linear regime, such a wave packet is nearly completely reflected so
that the interference of the incident and reflected waves creates a standing
wave pattern in front of the reflection point. 
Far outside the lattice, the period of the pattern is fixed by the
wave vector $k_0$ of the incident wave packet and is therefore 
of the order of the lattice period $\ell_{\rm Br}$.
Inside the lattice, the standing wave pattern originates instead from 
the interference of a forward and a backward Bloch waves: 
the amplitude of the fast oscillations follows from the density
modulation of each Bloch wave function with a periodicity $\ell_{\rm
Br}$, while the local period of the slower modulation which originates from
the interference of the two Bloch waves $k=k_{\rm Br}\pm\Delta k$ is equal
to $2\pi/\Delta k$.
As we approach the reflection point, the Bloch waves approach to the
band edge, so that $\Delta k\rightarrow 0$ and the period of the
modulation is strongly increased with respect to the lattice period,
although it still remains much shorter than the size of incident wave; 
this effect is apparent in the leftmost snapshot of fig.\ref{fig:Solit}.

Provided the nonlinear term is sufficiently small, the interactions do
not wash out the standing wave interference pattern~\cite{Lincoln} but simply
blue-shift the local band edge at the spatial position of the
rightmost antinodes; in this way, the wave packet frequency is pushed
out from the local gap and valence band states result available for
propagation.
The resulting short pulses, stabilized by the effective
attractive interaction, can therefore propagate along the lattice as solitonic
objects.
The higher the density, the larger the number of pulses for which this
mechanism is effective and which are then able to propagate along all
the lattice without being reflected.
By carefully choosing the density, a single soliton can be launched
along the lattice (fig.\ref{fig:Solit}); unfortunately, given the complexity
of the nucleation process, it is not physically evident how the
parameters of the soliton (e.g. the group velocity $v_g$ and the
peak density) can be controlled by acting on the parameters of the wide
incident Bose-condensate.

\begin{figure}[htbp]
\includegraphics[width=3in]{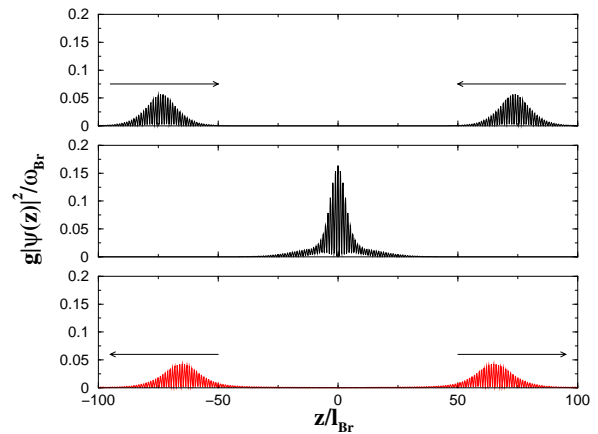}
\caption{
Collision process between a pair of gap solitons generated at either
end of a lattice with the same parameters as in figs.\ref{fig:Solit}
and propagating along the lattice with opposite velocities. All the
collisional dynamics takes place in the region $|z/\ell_{\rm Br}|<100$
where the lattice profile is flat.
\label{fig:Collis}}
\end{figure}

Once the pulse has crossed the flat region of the lattice and
has got to its opposite end, the effective mass of the atoms 
becomes positive again and the pulse is
immediately broadened under the combined effect of group velocity
dispersion and mean field repulsion (see the two last snapshots in fig.\ref{fig:Solit}). 
A proof of the solitonic nature of the generated pulse is obtained from a study
of its collisional dynamics (fig.\ref{fig:Collis}): a pair of such pulses symmetrically 
generated at the two ends of the lattice collide in the middle of the
lattice. Their solitonic properties result clearly from the fact that
their shapes as well as the number of atoms contained in each of them are only weakly affected by the
collision process.
The small broadening of the pulses that can be observed in
fig.\ref{fig:Collis} is a signature of the fact that the nonlinear
wave equation in a periodic potential is integrable only in an approximate 
way~\cite{AcevesWabnitz,SterkeNLSE}.

Thanks to the effective mass $m_{\rm eff}$ much smaller than the free
space mass $m_0$ (for the parameters of fig.\ref{fig:Solit} $m_{\rm
eff}=0.07\,m_0$), the solitonic width is significantly larger than the free
space healing length at the same value of the density; 
for a typical value of the lattice period of the order of
$0.5\,\mu\textrm{m}$, the solitonic length turns out of the order of
$10\,\mu\textrm{m}$ which is well within the capabilities of actual detection
systems. 
The mean-field interaction energies required for
the observation of solitonic effects are in general smaller or of the
order of one tenth of the recoil energy $\omega_{\rm Br}$: 
for the most relevant case of $^{87}$Rb atoms and $^{23}$Na, this
value corresponds to reasonable densities of the order of
$10^{14}\;\textrm{cm}^{-3}$.
The characteristic time for the modulational instability and soliton
formation processes described in the previous sections 
is of the order of $\omega_{\rm Br} t\simeq 500$, which means $t\simeq
20\,\textrm{ms}$ for $^{23}$Rb atoms and $t\simeq 3\,\textrm{ms}$ for
$^{87}$Na.

The method here described for the generation of bright gap solitons is
significantly different from previous proposals~\cite{BECGapSolit}:
in our approach the soliton pulse shape originates not from the whole
BEC cloud, but only from the much shorter density bump corresponding
to an antinode of the standing matter wave. This fact allows one to
obtain short solitons from a wide BEC without the need for a dramatic
pulse compression under the effect of effectively attractive interactions.

\section{Gap solitons: a simple analytical model}
\label{sec:SolitAnal}
Provided the gap soliton is wide enough, only a narrow group of Bloch
states around a central wave vector $k_{\rm sol}$ are populated and an
accurate description can be analytically obtained within the so-called envelope
function framework: denoting with $u_{k_{\rm sol}}(z)$ the Bloch eigenfunction
at $k=k_{\rm sol}$, we write the wave function $\psi(z,t)$ of the
coherent matter pulse as the 
product of the slowly varying envelope
${\bar \psi}(z,t)$ and the fastly oscillating Bloch eigenfunction
$u_{k_{\rm sol}}(z)$
\begin{equation}
\eqname{EffMassApproxWF}
\psi(z,t)={\bar \psi}(z,t)\,u_{k_{\rm sol}}(z);
\end{equation}
in the following, the Bloch eigenfunction $u_k(z)$ is assumed to be
normalized according to
\begin{equation}
\frac{1}{\ell_{\rm Br}}\int_{0}^{\ell_{\rm Br}} \!dz\,|u_k(z)|^2=1,
\end{equation}
which corresponds to 
\begin{equation}
|a_f|^2+|a_b|^2=1.
\end{equation}
in the $({\rm f}, {\rm b})$ basis of \eq{TwoModes}.
If its variations of the envelope ${\bar \psi}(z,t)$ are slow enough, 
it can be shown~\cite{SterkeNLSE} that ${\bar \psi}(z,t)$ obeys a
simple integrable nonlinear Schr\"odinger equation (NLSE)
\begin{equation}
i\hbar\frac{\partial {\bar \psi}(z,t)}{\partial
t}=\left(-\frac{\hbar^2}{2m_{\rm eff}}\frac{\partial^2}{\partial
z^2}+g_{\rm eff}|{\bar \psi}(z,t)|^2\right){\bar \psi}(z,t);
\eqname{NLSE}
\end{equation}
in which the effective mass
$m_{\rm eff}$ and the effective coupling $g_{\rm eff}$ depend on the
central wave vector $k_{\rm sol}$; obviously, the stability of the
solitonic pulse requires $m_{\rm eff}<0$, i.e. $k_{\rm sol}$ close to
the upper edge of the valence band at $k_{\rm Br}$.
While an explicit expression for $m_{\rm eff}(k)$ has been already
given in \eq{meff}, the effective coupling $g_{\rm eff}(k)$ turns out
to be expressed in terms of the Bloch wave function by
\begin{equation}
\eqname{geff}
g_{\rm eff}(k)=\frac{1}{\ell_{\rm Br}} \int_{0}^{\ell_{\rm Br}}\!dz\,g_{\rm 1D}\,|u_k(z)|^4;
\end{equation}
within the two-mode ansatz \eq{TwoModes}, this quantity can be
rewritten in the simple form
\begin{equation}
\eqname{geff2modes}
g_{\rm eff}(k)=g_{\rm 1D}\left( |a_{\rm f}|^4+|a_{\rm b}|^4+4|a_{\rm f}|^2 |a_{\rm b}|^2\right)
\end{equation}
where $a_{{\rm f},{\rm b}}$ are the
projections of the Bloch eigenfunction on the forward and backward
propagating waves.
Notice that the density modulation of the Bloch wavefunction at gap edge
($|a_{\rm f}|^2=|a_{\rm b}|^2=1/2$) makes the effective coupling a factor
$3/2$ larger than the one far from the gap, i.e. the free-space one.

\begin{figure}[htbp]
\includegraphics[width=3in]{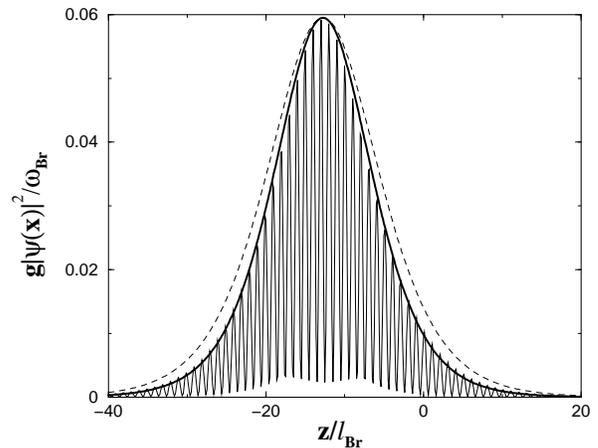}
\caption{
Comparison of the solitonic pulse at $\omega_{\rm Br}t=1700$ in 
fig.\ref{fig:Solit} with the analytical prediction \eq{solitAnal} for
the envelope (solid line). Approximate prediction obtained using the
band-edge values ($k_{\rm sol}=k_{\rm Br}$ for $g_{\rm eff}$ and
$m_{\rm eff}$.
\label{fig:Solit2}}
\end{figure}

Under these assumptions, the envelope ${\bar \psi}_{\rm sol}$ of the
solitonic wave packet has the simple expression~\cite{SolitReview}
\begin{equation}
{\bar \psi}_{\rm sol}(z)={\bar \psi}_{\rm
max}\,\textrm{sech}\left(\frac{z-v_g t}{\xi_{\rm sol}}\right)
\eqname{solitAnal}
\end{equation}
with the width $\xi_{\rm sol}$ given by
\begin{equation}
\eqname{xisol}
\xi_{\rm sol}=\sqrt{\frac{\hbar^2}{m_{\rm eff}g_{\rm eff}|{\bar \psi}_{\rm max}|^2}};
\end{equation}
as expected, the size $\xi_{\rm sol}$ of the soliton is of the order
of the healing length
$\xi=\hbar/\sqrt{2 m_{\rm eff}g_{\rm eff}|{\bar \psi}_{\rm max}|^2}$.
The accuracy of this approximate description is apparent in
fig.\ref{fig:Solit2} where we compare the numerically obtained wave packet
with the analytical prediction \eq{solitAnal} for the envelope; the central
wave vector $k_{\rm sol}$ of the wave packet has been determined from the group
velocity by means of \eq{vg}, the envelope amplitude ${\bar
\psi}_{\rm max}$ from the peak density of the pulse.

\section{Conclusions}
\label{sec:Conclu}

In this paper we have theoretically investigated the transmission dynamics of
coherent matter pulses (such as those that can be extracted from
Bose-Einstein condensates) which incide on finite optical lattices;
such systems are matter wave
analogs of the nonlinear Bragg fibers currently studied in nonlinear
optics.

At linear regime (i.e. in the low-density or weak-interaction limit), we have characterized the dependence of the
intensity and shape of the transmitted pulse on the velocity and size of the
incident condensate in terms of the dispersion of matter waves inside
the lattice; as in the case of light waves in periodic dielectric
structures or in the case of electrons in crystalline solids, the dispersion of
matter waves in the periodic potential of optical lattices is
in fact characterized by allowed bands and forbidden gaps.

The dynamics in the presence of interactions is found to be
even richer: an interpretation of the numerically predicted effects is
put forward in terms of familiar concepts from nonlinear optics, such
as optical limiting, optical bistability, and modulational
instability.

In particular, we have investigated a possible way of generating narrow bright
gap solitons from a wide incident Bose-condensate: the
modulational instability is seeded from the strongly modulated density profile
of the standing wave which is formed  in front of the
finite optical lattice by the interference of the incident and
reflected matter waves.
The solitonic nature of the generated pulses has been checked from
their shape which is in excellent agreement with a simple
analytical model as well as from their dynamical and collisional properties.

Finally, we have verified that the range of physical parameters that
is required for the observation of the effects predicted in the
present paper falls well within the possibilities of actual
experimental setups.

\begin{acknowledgments}
I.C. acknowledges financial support from the EU (``Marie Curie''
Fellowship of programme "Improving Human Research Potential and the
Socio-economic Knowledge Base" under contract number HPMF-CT-2000-00901).
I.C. is grateful to Yvan Castin, Lincoln Carr and Anna Minguzzi for
useful discussions. Laboratoire Kastler Brossel is a Unit\'e de
Recherche de l'\'Ecole Normale Sup\'erieure et de l'Universit\'e Paris
6, associ\'ee au CNRS.

\end{acknowledgments}


\begin{thebibliography}{99}

\bibitem{ColdAtomsLatticeExp} C. S. Adams and E. Riis, Prog. Quantum
Electron. {\bf 21}, 1 (1997); E. Peik, M. Ben Dahan, I. Bouchoule,
Y. Castin, and C. Salomon Phys. Rev. A 55, 2989-3001 (1997);
Q. Niu, X.-G. Zhao, G. A. Georgakis, and M. G. Raizen,
Phys. Rev. Lett. {\bf 76}, 4504 (1996); S. R. Wilkinson,
C. F. Bharucha, K. W. Madison, Q. Niu, and M. G. Raizen,
Phys. Rev. Lett. {\bf 76}, 4512 (1996)

\bibitem{BEC} {\em Bose-Einstein condensation in atomic gases},
Proceedings of the International School of Physics Enrico Fermi,
Course CXL, edited by M. Inguscio, S. Stringari, and C. Wieman (IOS
Press, Amsterdam, 1999).

\bibitem{BECLatticeExp} B. P. Anderson and M. Kasevich, Science {\bf
282}, 1686 (1998); S. Burger {\em et al.}, Phys. Rev. Lett. {\bf
86}, 4447 (2001); O. Morsch, J. H. M\"uller, M. Cristiani,
D. Ciampini, and E. Arimondo, Phys. Rev. Lett. {\bf 87}, 140402
(2001); 
M. Greiner, I. Bloch, O. Mandel, T. W. H\"ansch, and T. Esslinger,
Phys. Rev. Lett. {\bf 87}, 160405 (2001)


\bibitem{BECGapSolit} O. Zobay, S. P\"otting, P. Meystre, and
E. M. Wright, Phys. Rev. A {\bf 59}, 643 (1999);  S. P\"otting,
O. Zobay, P. Meystre, E.M. Wright, J. Mod. Opt. {\bf 47}, 2653 (2000);
A. Trombettoni and A. Smerzi, Phys. Rev. Lett. {\bf 86}, 2353 (2001)

\bibitem{InstabilityBEC} B. Wu and Q. Niu, Phys. Rev. A {\bf 64},
061603 (2001); V. V. Konotop and M. Salerno, cond-mat/0106228 (2001);
F. Kh. Abdullaev {\em et al.}, cond-mat/0106042 (2001)

\bibitem{PBG} E. Yablonovitch, J.Phys.: Condens.Matter {\bf 5}, 2443
(1993); E. Burstein and C. Weisbuch, eds., {\em
Confined electrons and photons}, Plenum Press, New York, 1995.

\bibitem{SterkeReview} C. M. de Sterke and J. E. Sipe in {\em Progress
in Optics} vol.XXXIII, ed. E. Wolf, Elsevier Science, Amsterdam, 1994,
pag. 203

\bibitem{BECLatticeTh} K. Berg-Sorensen and K. Molmer, Phys. Rev. A
{\bf 58}, 1480 (1998); D. Jaksch, C. Bruder, J. I. Cirac,
C. W. Gardiner, and P. Zoller, Phys. Rev. Lett. {\bf
81}, 3108 (1998);  S. Potting, M. Cramer, C. H. Schwalb, H. Pu,
P. Meystre, Phys. Rev. A {\bf 64}, 023604 (2001)

\bibitem{GapSolitExp} B. J. Eggleton, R. E. Slusher, C. M. de Sterke,
P. A. Krug, and J. E. Sipe, Phys. Rev. Lett. {\bf 76}, 1627 (1996);
B. J. Eggleton, C. M. de Sterke, and R. E. Slusher, J. Opt. Soc. Am. B
{\bf 16}, 587 (1999); S. Pitois, M. Haelterman, and G. Millot,
Opt. Lett, {\bf 26}, 780 (2001)

\bibitem{SelfPulsing} C. M. de Sterke and J. E. Sipe, Phys. Rev. A
{\bf 42}, 2858 (1990); C. M. de Sterke, Phys. Rev. A {\bf 45}, 8252 (1992); H. G. Winful, R. Zamir and S. Feldman,
Appl. Phys. Lett. {\bf 58}, 1001 (1991); C. M. de Sterke, Phys. Rev. A
{\bf 45}, 8252 (1992); A. B. Aceves, C. de Angelis, and S. Wabnitz,
Opt. Lett. {\bf 17}, 1566 (1992); N. M. Litchinitser, G. P. Agrawal,
B. J. Eggleton, and G. Lenz, Optics Express {\bf 3}, 411 (1998)

\bibitem{NLO}  R. W. Boyd, {\it Nonlinear Optics}, Academic Press,
London, 1992; P. N. Butcher and D. Cotter, {\em The elements of
nonlinear optics}, Cambridge University Press, Cambridge, 1993

\bibitem{Ozeri} N. Friedman, R. Ozeri, and N. Davidson,
J. Opt. Soc. Am. B {\bf 15}, 1749 (1998)

\bibitem{Santos} L. Santos and L. Roso, Phys. Rev. A {\bf 57}, 432 (1998)

\bibitem{Santos2} L. Santos and L. Roso, Phys. Rev. A {\bf 58}, 2407
(1998)

\bibitem{Santos3} L. Santos and L. Roso, Phys. Rev. A {\bf 60}, 2312 (1999)

\bibitem{AtomFP} I. Carusotto and G. C. La Rocca,
Phys. Rev. Lett. {\bf 84}, 399 (2000); I.Carusotto and G.C.La Rocca in
{\em Bose-Einstein condensates and atom lasers} eds. S. Martellucci,
A. N. Chester, A. Aspect, and M. Inguscio (Kluwer Academic/Plenum
Publishers, New York, 2000), pag.153

\bibitem{waveguide} K.Bongs {\em et al.}, Phys. Rev. A {\bf 63},
031602 (2001); A. G\"orlitz {\em et al.}, Phys. Rev. Lett. {\bf 87},
130402 (2001); F. Schreck {\em et al.}, Phys. Rev. Lett. {\bf 87},
080403 (2001)

\bibitem{BECSolitTh} W. P. Reinhardt and C. W. Clark, J. Phys. B {\bf
30}, L785-L789 (1997); A. D. Jackson, G. M. Kavoulakis, and C. J. Pethick,
Phys. Rev. A {\bf 58}, 2417 (1998); Th. Busch and J. R. Anglin,
Phys. Rev. Lett. {\bf 84}, 2298 (2000); Th. Busch and J. R. Anglin, 
Phys. Rev. Lett. {\bf 87}, 010401 (2001)

\bibitem{SolidState} N. W. Ashcroft and N. D. Mermin, {\em Solid state
physics}, Saunders College Publishing, Orlando, 1976

\bibitem{CCT4} see exercise 13 in C. Cohen-Tannoudji, J. Dupont-Roc
and G. Grynberg, {\it Processus d'interaction entre photons et atomes}
(InterEditions/Editions du CNRS, Paris, 1988).

\bibitem{DBRMicrocav} {\em Confined photons and electrons}
eds. E. Burstein and C. Weisbuch (Plenum Press, New York, 1995).

\bibitem{ModInstSuppr} P. A. Ruprecht, M. J. Holland, K. Burnett, and
M. Edwards, Phys. Rev. A {\bf 51}, 4704 (1995); Yu. Kagan,
G. V. Shlyapnikov, and J. T. M. Walraven, Phys. Rev. Lett. {\bf 76},
2670 (1996)

\bibitem{AttractiveBEC} C. A. Sackett, H. T. C. Stoof, and
R. G. Hulet, Phys. Rev. Lett. {\bf 80}, 2031 (1998)


\bibitem{SolitAdiab} Y. S. Kivshar and B. A. Malomed,
Rev. Mod. Phys. {\bf 61} 763 (1989) and Rev. Mod. Phys. {\bf 63}, 211
(1991); E. N. Tsoy and C. M. de Sterke, J. Opt. Soc. Am. B {\bf 18}, 1
(2001).

\bibitem{SolitReview} J. R. Taylor, ed., {\em Optical Solitons}, 
Cambridge University Press, Cambridge, 1992

\bibitem{PhotInter} I. H. Deutsch, R. Y. Chiao, and J. C. Garrison,
Phys. Rev. A {\bf 47}, 3330 (1993)

\bibitem{AcevesWabnitz} A. B. Aceves and S. Wabnitz, Phys. Lett. A
{\bf 141}, 37 (1989)

\bibitem{BECSolitExp} S. Burger {\em et al.}, Phys. Rev. Lett. 83,
5198 (1999); J. Denschlag {\em et al.}, Science {\bf 287}, 97 (2000)

\bibitem{Lincoln} L. D. Carr, C. W. Clark, and W. P. Reinhardt,
Phys. Rev. A {\bf 62}, 063610 (2000); W.-M. Liu, B. Wu, and Q. Niu,
Phys. Rev. Lett. {\bf 84}, 2294 (2000)

\bibitem{SterkeNLSE} C. M. de Sterke and J. E. Sipe, Phys. Rev. A
{\bf 42}, 550 (1990); C. M. de Sterke and B. J. Eggleton,
Phys. Rev. E {\bf 59}, 1267 (1999)

\end{thebibliography}
\end{document}